\documentclass[12pt,a4paper]{revtex4}
\begin{document}
\title{Unimodular relativity and cosmological constant : Comments}
\author{ S.C. Tiwari}
\affiliation{Institute of Natural  Philosophy \\
C/o 1 Kusum Kutir, Mahamanapuri \\ 
Varanasi 221005, India }

\begin{abstract}
We show that the conclusion that matter stress-energy tensor
satisfies the usual covariant continuity law, and the cosmological constant is
still a constant of integration arrived at by Finkelstein et al (42, 340,
2001) is not valid.
\end{abstract} 	
\maketitle

A recent paper \cite{1} reports the status of dark matter
from the view point of  observational cosmology. It is also noted that
anisotropic cosmic microwave background radiation provides 'compelling
argument for a non-zero cosmological constant'. Theoretically, cosmological
constant was first introduced by Einstein, but later he felt that this term
reduced the 'logical simplicity' of the theory \cite{2}. The interested reader
may find extensive references to the literature in the reviews \cite{3}. In this
brief comment, we re-examine the problem of energy-momentum conservation law
in unimodular gravity. 	Anderson and Finkelstein \cite{4} envisage a cellular 
structure
of space-time, and develop unimodular theory of relativity based on an action
principle for a measure manifold with a fundamental measure $\mu(x)$.  The
unimodular condition is

\begin{equation}
\sqrt{-g} d^4 x = \mu(x) d^4 x
\end{equation}

To derive the field equations based on the action principle, the measure   
is assumed to be a fixed nondynamical field. Choosing $\mu(x)=1$ may give rise
to a simpler unimodular coordinate choice. The unimodular condition (1) is
incorporated in the action function using Lagrange's undetermined multiplier,
$\lambda(x)$ . The field equations derived from the variational principle
admit a cosmological term, and the cosmological constant is an integration
constant. In \cite{5} we point out that without an additional assumption on the
covariant divergence of the matter energy-momentum tensor. $T^{\mu\nu} $,   
the cosmological constant is not 'constant'. In a recent paper, Finkelstein
et al \cite{6} elucidate unimodular relativity emphasizing the role of a 
conformal
metric tensor $f_{\mu\nu}-$ 'the sole gravitational variable of unimodular
relativity'. We believe the approach based on $f_{\mu\nu}$ may have interesting
physics, specially due to the possibility of exploring Weyl geometry in this
framework. Here we focus our attention on the consequences of ambiguous
extended action, S' presented in Sec. III of \cite{6}. We use the notations 
of the
authors \cite{6}, and give main steps in the following. The matter Lagrangian
density is 	
\begin{equation}
L'_M = L_M + \Delta_ML						
\end{equation}
The ambiguity is assumed to be of the form
\begin{equation}
\Delta_ML = \left[\frac{\mu(x)}{\sqrt{-g}}-1 \right] l_M			
\end{equation}
where  $l_M$ is a function of matter field variables and $g_{\mu\nu}$.
The Lagrange multiplier $\lambda(x)$ is introduced to incorporate the
unimodular condition (1), and the action integral S' is constructed as usual,
see eqn. (13) in \cite{6}. Varying $g_{\mu\nu}$   gives the field equation
\begin{equation}
G^{\mu\nu}-\frac{\lambda}{2}g^{\mu\nu} = 8\pi GT'^{\mu\nu}
\end{equation}
Here $G^{\mu\nu}$ is the Einstein tensor, and the ambiguous energy-momentum 
tensor is
\begin{equation}
T'^{\mu\nu} = T^{\mu\nu} + \frac{1}{\sqrt{-g}}
\frac{\delta(\sqrt{-g}\Delta_ML)}{\delta g_{\mu\nu}}  	  
\end{equation}
It is straightforward to calculate $\lambda$ taking trace of (4)
\begin{equation}
\lambda = -\frac{8\pi GT' + R}{2}
\end{equation}
The field equation (4) reduces to
\begin{equation}
R_{\mu\nu} - \frac{R g_{\mu\nu}}{4} = 8\pi G (T'_{\mu\nu} - 
\frac{g_{\mu\nu}T'}{4})
\end{equation}
If the ambiguity in energy-momentum tensor is calculated following the
prescription (3), we get
\begin{equation}
\Delta T^{\mu\nu} = \frac{g^{\mu\nu}l_M}{2}			
\end{equation}
Evidently there is an error in eqns. (24) and (25) of \cite{6}. Using (8), and
taking trace of eqn. (4), we obtain
\begin{equation}
2\lambda = -R-8\pi G(T+2l_M)
\end{equation}
Both $\lambda$   and  $l_M$ get eliminated in the final field equation
\begin{equation}
R_{\mu\nu} - \frac{R g_{\mu\nu}}{4} = 8\pi G(T_{\mu\nu}-\frac{g_{\mu\nu}T}{4})
\end{equation}
Finkelstein et al \cite{6} note that $T'_{\mu\nu}$ is not covariantly
continuous in unimodular relativity and state that that seems to justify 
modified covariant divergence law, eqn(7) of \cite{5}. However, the 
authors assert that $T_{\mu\nu}$ satisfies the usual covariant divergence 
law, and the cosmological constant is still a constant of integration. 
Though authors do not state it explicitly,the additive ambiguity in 
$T_{\mu\nu}$, and the discussion following eqn(23) in their paper seem 
to imply that somehow the ambiguity in the matter field Lagriangian leads 
to this result. Does this result that usual covariant continuity law 
holds for $T_{\mu\nu}$ follow from their theory? 

To analyze this question first we make a remark based on \cite{5}. 
In the notation of that paper eqn(1) and eqn(7) of \cite{5} give L to 
be a constant using the Bianchi identity. From eqn(6) of that paper it 
follows that R+T is constant. One could assume the constancy of R+T 
and infer the covarint continuity of the energy-momentum tensor. 
Here also we use the Bianchi identity and derive the covariant divergence 
law for $T{\mu\nu}$ in the Finkelstein et al theory. Taking covariant 
divergence of eqn(4) above we get
\begin{equation}
8\pi G {T^{'\mu\nu}}_{:\nu} = -\frac{1}{2} g^{\mu\nu} \lambda_{\,\nu}
\end{equation}
From eqn(5), it is straightforward to calculate 
\begin{equation}
8\pi G {T^{\mu\nu}}_{:\nu} = -\frac{1}{2}g^{\mu\nu} (\lambda + 
8\pi G l_M)_{,\nu}
\end{equation}
Evidently $T_{\mu\nu}$ is not covariantly continuous as shown by eqn(12). 
Though $l_M$ does not appear in the field eqns (7) and (10),from 
expression(9) as well as eqn(12) it becomes clear that $l_M$ mofifies 
the cosmological constant term. Let us denote it by 
$\lambda_{eff}(=\lambda+8\pi Gl_M)$.

If the consistency of the theory given by Finkelstein et al has to be 
maintained then following conclusions are inevitable:
\begin{enumerate}
\item  If we impose the condition that the covariant divergence of 
$T_{\mu\nu}$ vanishes then $\lambda_{eff}$ is constant. From eqn(9) it 
follows that $R+8\pi GT$ is also constant. However $T{'\mu\nu}$ still 
satisfies the modified covariant divergence law, eqn(11) and $\lambda$ 
is not necessarily a constant. Imposing the further condition that 
covariant divergence of $T'_{\mu\nu}$ is also zero, $\lambda$ too becomes 
a constant. As a consequence $R+ \pi GT'$ as well as $l_M$ also become constant.
\item
Alternatively, imposing the condition that $R+8\pi GT$ and $R+8\pi GT'$ are 
constant,the constancy of  $\lambda$ and $l_M$ as well as the covariant 
continuity of both primed and unprimed energy-momentum tensors follow.
\item
The principal result of \cite{6}  therefore is an assumption, not a 
consequence of the theory.
\end{enumerate}

We may point out that it is also inconsistent to allow cosmological constant 
to be a field variable and simultaneously require covariant continuity of 
energy-momentum tensor as is done by Ng and van Dam \cite{7}. To end the 
paper, let us first note a mathematical result:both $\lambda$ and $l_M$ are 
variable such that the sum $\lambda+8\pi Gl_M$ is required to be zero. In that case eqn(4) 
reduces to the standard Einstein field equation without cosmological constant. 
Though the physical significance of ambiguity introduced in \cite{6} is not 
clear, it seems interesting to speculate that $\lambda$ and $l_M$ in eqn(9) 
correspond to geometric and vacuum energy aspects of the unimodular world. 
Does that mean that in the standard theory with Einstein field equation both 
contributions cancel each other under this condition ? We leave the answer to this question 
for future investigations. 

I am grateful to Profs. D. Finkelstein and A.A.Galiutdinov for extensive 
e-correspondence on this subject. Library facility of Banaras Hindu 
University, Varanasi is acknowledged.

\end{document}